\documentclass[aps, 10pt]{revtex4}
\usepackage{graphicx,subfigure}
\usepackage{epsfig}
\usepackage{amsmath}
\usepackage{amsfonts}
\usepackage{amssymb}
\begin{document}
\title{ Random Matrices and Quantum Hamilton-Jacobi Method}
\author{ K. Haritha and K.V.S.Shiv Chaitanya $^1$}
\email[]{ chaitanya@hyderabad.bits-pilani.ac.in}
\affiliation{$^1$Department of Physics, BITS Pilani, Hyderabad Campus, Jawahar Nagar, \\Shamirpet Mandal,
Hyderabad, India 500 078.}

\begin{abstract}
In this paper, we start with the quantum Hamilton-Jacobi approach and show that the underlying complex pole evolution of the Schr\"odinger equation is described by the quantum action in terms of a  random matrix. The wave function is given by the random matrix probability distribution function. In literature this is known as the famous Cole-Hopf Transformation. 
\end{abstract}
\maketitle
\section{Introduction}
The classical action first appeared in the quantum mechanics as quantization rule given by  Wilson and Sommerfeld \cite{ws}, wherein the energy levels are approximately obtained by setting the classical action variable equal to an integral multiple of the Planck's constant. Later on several attempts were made to construct a quantum analogue to the classical action. Defining a quantum action allows us to transform one set of canonically conjugate coordinates and momenta to another set, analogous to the classical mechanics. This is known as quantum transformation, originated from the works of Jordon \cite{jor} and Dirac \cite{dir}. 
Dirac  defined the following quantum action $S$ 
as the quantum analogue  of the classical action function \cite {dir1, dir2} such that
\begin{equation}
(\xi'|\alpha')= exp(iS/\hbar)\label{dirk}
\end{equation}
where  $(\xi
'|\alpha')$ is the transformation function that connects the $\xi$ representation to $\alpha$ representation. Starting from the definition (\ref{dirk}), he has shown that the quantum transformation equations have the same form as the classical equations. 

Dirac quantum action $S$ has connections to the Feynmann path integral formulation of quantum mechanics \cite{fey,schw,fin}. The comprehensive examination of the quantum transformation theory is done by Schwinger in the references \cite{schw, fin, schw1}. Schwinger postulated the action principle such that one obtains operators $\delta S$ by the variation of one operator $S$ \cite{fin, schw1}. He derived the Schr\"odinger equation from an "action" principle in a way very similar to the Hamilton derivation of the Hamilton-Jacobi equation from the classical action principle \cite{schw1}. It is shown in the references \cite{fin, schw1} that the Schwinger's action principle is identical to the Dirac's original expression. 

 Leacock and Padgett developed quantum transformation theory as a generalization of classical Hamilton-Jacobi theory \cite{qhj,qhj1}.  In the Quantum Hamilton-Jacobi approach, we start with the Schr\"odinger equation,
\begin{equation}
- \frac{\hbar^2}{2m}\frac{\partial^2 }{\partial x^2}\psi(x)+ V(x) \psi(x) = E
\psi(x). \label{sc} 
\end{equation}
We define a function in terms of quantum action or quantum characteristic function $S$ analogous to the classical characteristic function by the relation 
\begin{equation}
\psi= exp \left[\frac{iS} {\hbar}\right].   \label{ac}
\end{equation}
The equation  (\ref{ac}) is known as Cole-Hopf transformation and when substituted in (\ref{sc}), gives
\begin{equation}
\left(\frac{d}{dx}S\right)^2 -i \hbar \frac{d}{dx}\left(\frac{d}{dx}S\right) = 2m (E
- V(x)). \label{qhj0} 
\end{equation}
The quantum momentum function $p$ is 
\begin{equation}
p =\frac{d}{dx} S. \label{mp}
\end{equation}
Substituting (\ref{mp}) in (\ref{qhj0}) gives the Quantum Hamilton-Jacobi equation
for $p$ as 
\begin{equation}
p^2 - i \hbar \frac{dp}{dx} = 2m (E - V(x)), \label{qhj1}
\end{equation}
which is the Riccati equation.  It should be clear to the readers from the equation (\ref{qhj1}) when $\hbar\rightarrow 0$ we get the classical momentum function $p_c^2= 2m (E - V(x))$. The relation between the quantum momentum function and the wave function is given by
\begin{equation}
p=i\hbar\frac d {dx} ln \psi(x).  \label{lg1}
\end{equation}
It is shown by Leacock and Padgett \cite{qhj, qhj1} that the action angle variable gives rise to the exact quantization condition
\begin{equation}
J(E) \equiv  \frac{1}{2\pi} \oint_C{pdx} = n\hbar.       \label{act}
\end{equation}
We would like to point out that the transformation (\ref{ac}) is used in the Wentzel, Kramers, Brillouin (WKB) approximation and is the semi-classical approach. The $S$ defined in WKB is the classical action whereas $S$ defined in equation (\ref{ac}) is the quantum action. For more details refer to appendix C of \cite{qhj1}. From the equation (\ref{qhj1}) it is evident that the term $i \hbar \frac{dp}{dx}$ is the one that gives rise to quantum effects and is absent in classical momentum function. Hence, the $S$ defined in equation (\ref{ac}) is the effective action or the quantum action \cite{zo}. The connection  between the QHJ and Dirac's action is given in the appendix. Readers should note that all the definitions of quantum action defined in literature have correspondence to the Dirac's action (\ref{dirk}).

In this paper, we show that the quantum action $S$ is described by the random matrix of Gaussian unitary ensemble and the Cole-Hopf transformation (\ref{ac}) is the random matrix probability distribution function. We use the fact that quantum Hamilton-Jacobi can be solved in two different ways. One way of  solving is by using the Riccati equation which leads to the action angle variable gives rise to the exact quantization condition (\ref{act}). The other way is using the Burger equation. We show that this way of solving quantum Hamilton-Jacobi gives quantum action.  In literature, the transformation (\ref{ac}) is used to map the Schr\"odinger equation to the Burger equation \cite{chj}.  It is well known that underlying complex pole evolution of Burger equation in turn the Schr\"odinger equation is described by the Kirchhoff equations \cite{kvs1}
 \begin{eqnarray}
 \dot{x}_i=2\Gamma i\sum_{i\neq j}^n\frac{1}{(x_i-x_j)} +i\mathcal{W}(x_i),\label{kir14}
 \end{eqnarray}
 where $x_i$ is the position of charges and $\mathcal{W}(x_i)$ is the superpotential. These equations describe the evolution of $n$ point vortices in Hydrodynamics. The solution to the stationary Kirchhoff equations (\ref{kir14}), that is $\dot{x}_i=0$, is by found by the Stieltjes electrostatic model \cite{st,st1}. The Stieltjes electrostatic model enables the Langrangian formalism.  We show that when Stieltjes electrostatic model is used for solving quantum Hamiltonian-Jacobi it leads to quantum action $S$.
In random matrix theory the probability distribution function of random matrix is the stationary solution of Fokker-Planck equation. We observe that when Fokker-Planck equation is mapped to Schr\"odinger equation by Cole-Hopf transformation. We also use the fact that Stieltjes electrostatic model  is  used to solve the Dyson-log-gas model in random matrix theory \cite{met}. The equations obtained in Dyson-log-gas model are the Kirchhoff equations. In the reference \cite{iopj} the authors have shown that equations obtained in Dyson-log-gas model is obtained by minimizing action. Hence, the claim of this paper is the Dirac action is same is the action obtained in reference \cite{iopj}  by using the Stieltjes electrostatic model.
\section{Random Matrices}

In physics, random matrices' initial application was to describe the excited states of the atomic nuclei \cite{wig} by Wigner. Random matrix theory has a wide variety of applications from chiral symmetry breaking in quantum chromodynamics, quantum chaos to many-body physics  \cite{qcd, qtqd, sas,cs, phi,tar, zaf, guy, sud, dan}. Here the dynamics of the underlying system is treated as an ensemble of random matrix $H$ and is described by the probability distribution  of the eigenvalues $x_i$\cite{met}
 \begin{equation}\label{pdfo}.
 {\cal P}(x_1,\ldots,x_N)\,d\Lambda
 = C_ne^{-\beta Tr(H) }
 \,\,d\Lambda,
 \end{equation}
 here the index $\beta=1, 2, 4$ and \begin{eqnarray}
 Tr(H)=-\sum_{i=1}^NV(x_i)-
 \beta\sum_{i<j}^Nln\vert x_i-x_j\vert \label{hamp}
 \end{eqnarray}
 where $V(x_i)$ is the potential then $\prod_{i<j}\vert x_i-x_j\vert$ is the Vandermonde determinant, $d\Lambda=dx_1 \ldots dx_N$ and $c_n$ is the constant of proportionality. For details please refer to \cite{met, rm, ran1,ran2,ran3, ran4}. The $\beta$ in equation (\ref{pdfo}) characterizes the real parameters of the symmetry class of orthogonal, unitary and symplectic respectively for the random matrix $H$.

 As the random matrix is invariant under the symmetry group these ensembles are called the invariant ensembles. The probability distribution function can be rewritten as \cite{rm}
  \begin{equation}\label{pdf1}
 {\cal P}(x_1,\ldots,x_N)\,d\Lambda
 =c_n \prod_{i=1}^N w_\beta (x_i)
 \prod_{i<j}\vert x_i-x_j\vert^\beta
 \,\,d\Lambda
 \end{equation}
 where $w_\beta (x_i)$ is the weight function of the classical orthogonal polynomials.
 The classical orthogonal polynomials are classified into three different categories depending upon the range of the polynomials. The polynomials in the interval $(-\infty;\infty)$ with weight function $w=e^{-x^2}$ are the Hermite polynomials.
 In the interval $[0;\infty)$ with weight function $w=x^be^{-x}$ are the Laguerre polynomials. In the interval $[-1;1]$ with weight function $w=(1+x)^{a}(1-x)^b$ are the Jacobi polynomials.
 
 The random matrices are used to model the real world problems and one of the well known model is the Dyson-log-gas model. The probability distribution function for Dyson-log-gas model is given by \cite{met,dyson}
 \begin{equation}
 P(x_1x_2.....x_n)=c_ne^{-\beta H}=\frac 1 {c_\beta}e^{-\beta (\sum_{j=1}^{n}x_j^2+\sum_{j,k}ln(x_j-x_k))}. \label {tt}
 \end{equation}.
Dyson treated this as the partition function and the random matrix $H$ as potential energy.
 Then the thermodynamic equilibrium for this model is obtained by minimizing the energy 
 \begin{equation}
 \frac{d}{dx_j}H=  \sum_{1\leq j \leq k, j\neq k }^{n}\frac 1 {x_j-x_k}-x_j=0.\label{ssv}
 \end{equation}
 The equation (\ref{ssv}) is the system of $n$ linear equations which is solved using the Stieltjes electrostatic model \cite{met}. 
Theorem $1$ : \textit{In Stieltjes electrostatic model, the following Lagrangian}  
 \begin{eqnarray}
 L&=& -\sum_{1\leq i <k\leq n}^n ln\vert x_i-x_j\vert-p\sum_{i=1}^n ln(\vert 1-x_i\vert)-q\sum_{i=1}^n ln(\vert 1+x_i\vert),\label{st}
 \end{eqnarray}
\textit{is considered, where $x_i$ is the position of charges,  with interaction forces  for the $n$ moving unit charges arising from a logarithmic potential which are in between the two fixed charges $p$ and $q$ at $-1$ and $1$ respectively on a real line.} Then he proved in ref \cite{st, st1} that the expression (\ref{st}) becomes a minimum when $(x_1, x_2, \cdots,x_n)$ are the zeros of the Jacobi polynomial.
For the proof refer to Szego's book (section 6.7) \cite{sz}.
The zeros of the Laguerre and the Hermite polynomials admit the same interpretation. Thus, we solve (\ref{ssv}) using Stieltjes electrostatic model.
By using the following identity \cite{met}
\begin{eqnarray}
 \sum_{1\leq j \leq k, j\neq k }^{n}\frac 1 {x_j-x_k}=\frac{f''(x_j)}{2f'(x_j)}.\label{lhp},
\end{eqnarray}
and substituting in equation (\ref{ssv}) reduces to
\begin{equation}
f''(x_j)-2x_jf'(x_j)=0. \label{hui}
\end{equation}.
This system has thermodynamic equilibrium at the zeros of the Hermite orthogonal polynomial. Thus the differential equation (\ref{hui}) reduces to Hermite  differential equation
\begin{equation}
f''(x)-2x_jf'(x)+2\lambda  f(x)=0,
\end{equation}
and the solutions are the Hermite orthogonal polynomials 
\begin{equation}
f (x)=c_n e^{-{\frac {x^{2}}{2}}}\cdot H_{n}\left(x\right),\qquad n=0,1,2,\ldots .\label{hor}
\end{equation}
In the next section we present a connection between the random matrices and the self-adjoint operator.

 \section{Random matrices and the self-adjoint operator}
 It is well known that the probability distribution function of a random matrix is the stationary solution of the Fokker-Planck equation \cite{met}. Then  the the Fokker-Planck operator is mapped to the  Schr\"odinger equation is studied in reference  \cite{nelson}. We observe the mapping of Fokker-Planck operator  to  Schr\"odinger equation is through a  by Cole-Hopf transformation. 
 
 Theorem $2$: \textit{The probability distribution function in the equation (\ref{pdfo}) is a stationary solution of the Fokker-Planck equation} \cite{met}. 
 
 Proof: Differentiating the probability distribution function (\ref{pdfo}) gives
 \begin{equation}
 \frac{\partial {\cal P}}{\partial x_j} =-\beta P \frac{dH}{dx_j},\label{u}
 \end{equation}
where $H$ is random matrix. in general it need not be of the form (\ref{hamp}). 
 Differentiating the equation (\ref{u}) again w.r.t $x_j$,  
follows that 
 \begin{equation}
 \sum_{j=1}^N \frac{1}{\beta} \frac{\partial^2 {\cal P}}{\partial x_j^2}+\frac{\partial }{\partial x_j}(H'(x_j){\cal P})\}=0.
 \end{equation} 
 Since the Fokker-Planck equation is 
 \begin{equation}
 \frac{\partial P}{\partial t}= \sum_{j=1}^N\{1/\beta \frac{\partial^2 {\cal P}}{\partial x_j^2}+\frac{\partial }{\partial x_j}(H'(x_j){\cal P})\},
 \end{equation}
 it follows that ${\cal P}$ is a stationary solution of the Fokker-Planck equation. This ends the proof.

 Theorem $3$: \textit{The Fokker-Planck operator is self-adjoint if and only if the drift term is the gradient of a potential \cite{nelson}.} 
 
 Proof: We start with the Fokker-Planck equation 
 \begin{equation}\label{fok}
 \frac{\partial P}{\partial t}= \sum_{j=1}^N\{1/\beta \frac{\partial^2 {\cal P}}{\partial x_j^2}+\frac{\partial }{\partial x_j}(H'(x_j){\cal P})\},
 \end{equation}
By defining a new function
 \begin{equation}
 {\cal P}=Ne^{-\frac{\beta }{2}H(x)+\frac{1}{2}Et}\psi\label{uio}
 \end{equation}
 where $\beta$ is a parameter which will be identified later and $N$ is a constant. 
 Thus, the Fokker-Planck operator (\ref{fok}) reduces to
 \begin{eqnarray}
\frac{1}{\beta} \frac{\partial^2}{\partial x_j^2}\psi+   V\psi&=&\frac{1}{2}E\psi\label{yuo}
 \end{eqnarray}
 which can be interpreted as the Schr\"{o}dinger operator for appropriate $\beta$  with the potential defined as
 \begin{equation}
 V=\frac{1}{2}\frac{\partial^2 H(x_j)}{\partial x_j^2}-\frac{\beta}{4}(\frac{\partial H(x_j)}{\partial x_j})^2.\label{por}
 \end{equation}
where $\mathcal{W}(x)=\frac{\partial H}{\partial x_j}$ is superpotential defined in equation (\ref{gh}) and for $\beta=2$ the equation (\ref{yuo}) reduces to the Schr\"{o}dinger type equation. Hence, from equation (\ref{por}) the drift term is the gradient of the potential. This ends the proof. 

Corollary $1$: Fokker-Planck reduces to the  Schr\"{o}dinger equation (\ref{sc}) in imaginary time formalism that is $t\rightarrow iT$ in equation (\ref{fok}) and the  term $H(x)-\frac{E}{2}t$ in exponential (\ref{uio}) is the canonical transformation  (\ref{hp}).

Proof: When exponential factor $e^{i(H(x)-\frac{1}{2}ET)}$ that is $\beta=-2i$ and $i\mathcal{W}(x)=\frac{\partial H}{\partial x_j}$ Schr\"{o}dinger equation (\ref{yuo}) reduces to (\ref{sc}) with $V=-\frac{1}{2}(\mathcal{W}^2(x)-\mathcal{W}'(x))$. Thus, the term $H(x)-\frac{E}{2}T$ in exponential (\ref{uio}) is the Legendre transformation  (\ref{hp}). Then the quantum Hamilton-Jacobi connection to random matrix is immediate and in turn to Dirac's action. This ends proof.

 \section{Two  approaches of quantum Hamilton-Jacobi}
There are two ways of approaching quantum  Hamiltonian-Jacobi formalism. The first way is by using the quantum momentum function (\ref{mp}) and second way is through Stieltjes electrostatic model \cite{kvs}. In the first case, it is evident that the quantum momentum function is the solution to the Riccati equation (\ref{qhj1}). Thus, by analytically continuing $p$ to the complex plane and knowing the singularities of the quantum momentum function is sufficient to determine the wave functions \cite{sree}. The solutions of the Riccati equation have two types of singularities \cite{inc}, the fixed and the moving singularities. The fixed singular points of $p(x)$ are reflected in the potential and are energy independent. The pole structure of the Riccati equation is given in terms of quantum momentum function \cite{sree}
 \begin{equation}
 p=\sum_{j=1}^n\frac{-i\hbar}{x-x_j}+Q(x).\label{uf0}
 \end{equation}
  where $Q(x)$ is meromorphic function and $i\hbar$ are the residues of moving poles in the quantum momentum function \cite{sree}.
 The equation (\ref{uf0}) represents $n$ moving poles on a real line coming from  $\sum_{j=1}^n\frac{-i\hbar}{x-x_j}$ between the two fixed poles coming from the $Q(x_J)$.

  Theorem $4$: \textit{The equation (\ref{ufiu5}) admits orthogonal polynomial solutions and this leads to quantization provided the function $Q(x_j)$ is a superpotential.} \cite{kvs}
 
 Proof : Starting from equation (\ref{uf0}) and introducing the polynomial
 \begin{eqnarray}
 \psi(x)&=&(x-x_1)(x-x_2)\cdots (x-x_n)=\prod_{k=1}^n(x-x_j),\label{poly}
 \end{eqnarray}
 and substituting the equation (\ref{poly}) in the quantum momentum function (\ref{uf0}) and taking( $\hbar=1$) reads as 
 \begin{equation}
 p=- i\frac{f'(x)}{f(x)}+Q(x)\label{uf4}
 \end{equation}
 and substituting equation (\ref{uf4}) in equation (\ref{qhj1}) one gets
 \begin{eqnarray}
 -f''(x) + 2iQ(x)f'(x)+[Q^2(x)-iQ'(x)-E+V(x)]f(x)=0.\label{dif11}
 \end{eqnarray} 
 Finding the polynomial solutions to the equation (\ref{dif11}) leads to quantization. This is equivalent to demanding that $ [Q^2(x)-iQ'(x)-E + V(x)]$ is a constant. From equation (\ref{gh}) this is only possible if 
 $Q(x)=i\mathcal{W}(x)$ is the superpotential. Thus, the differential equation (\ref{dif11}) reduces to Sturm-Liouville equation. This ends the proof.
 
In the second case that is by Stieltjes electrostatic model, the problem is solved using the complex pole evolution of Burger's equation. 

Theorem $5$: \textit{The  Schr\"{o}dinger equation is mapped to Burger equation by the complex Cole-Hopf transformation} \cite{chj}.

Proof:  Substituting the complex Cole-Hopf transformation (\ref{ac}) in the Schr\"{o}dinger equation (\ref{sc}) gives the Riccati equation (\ref{qhj1}).
Differentiating the equation (\ref{qhj1}) with respect to $x$  gives
\begin{equation}
 -i \hbar \frac{\partial^2 p}{\partial x^2}-2p \frac{\partial p}{\partial x}+\frac{\partial V}{\partial x}=0\label{bugy}
\end{equation}

This is Burger's equation and $p$ is given in equation (\ref{mp}), this ends the proof. 

It is clear from the equation (\ref{bugy})  quantum momentum function $p$ is also solution to Burger's equation. Therefore, we consider the following pole structure quantum momentum function $p$ with $V(x)=0$, that is
 \begin{equation}
 p(x)=\sum_{k=1}^n\frac{-i\hbar}{x-x_k},\label{ufo1}
 \end{equation}
Theorem $6$: \textit{The complex pole expansions of the form (\ref{ufo1}) of the Burger's equation is satisfied if and only if  the following system of differential equations are satisfied or the complex poles evolve according to the system of $n$ linear equations given by}
 	\begin{equation}
 	\frac{d x_j}{dt}=\sum_{1\leq j\leq n,j\neq k}^n\frac{1}{x_k-x_j}.\label{ufie}
 	\end{equation}
 For proof please refer to \cite{chood}. The equation (\ref{ufie}) are known as Kirchhoff equation. Therefore, in the  stationary case $\frac{d x_j}{dt}=0$ the equation (\ref{ufie}), reduces to the free particle Kirchhoff equations (\ref{kir14}) for $\Gamma=\hbar/2$ and the superpotential $\mathcal{W}=0$.
 Thus by adding a potential term to  Kirchhoff equation (\ref{ufie}) and working in stationary case  reads as
 \begin{equation}
 \sum_{1\leq j\leq n,j\neq k}^n\frac{-i\hbar}{x_k-x_j}+Q(x_j)=0.\label{ufiu5}
 \end{equation}
 where $Q(x_j)$ is meromorphic function. If $x_j$ are eigenvalues, $i\hbar=1$ and $Q(x_j)=x_j$ then equation (\ref{ufiu5}) reduces to Dyson log-gas model (\ref{ssv}).
 
Theorem $7$ :  \textit{The stationary Kirchhoff equation (\ref{ufiu5}) are solved using Stieltjes electrostatic model and admits a classical polynomial solution if   $Q(x_j)$ is superpotential.} \cite{kvs1}

Proof: We start with the Kirchhoff equation (\ref{ufiu5}) and using (\ref{lhp}) we get
\begin{equation}
-f''(x_j)+2iQ(x_j)f'(x_j)=0. \label{huii}
\end{equation}
This system has thermodynamic equilibrium at the zeros of the classical orthogonal polynomial provided the  equation (\ref{huii}) reduces to the equation  (\ref{dif11}). Therefore, from theorem $4$ we get $Q(x_j)=i\mathcal{W}(x_j)$ is the superpotential. This ends the proof.
\section{Random matrices and Quantum Action}
It should be clear to the readers from  previous section that solving the quantum Hamiltonian-Jacobi through Riccati equation (\ref{qhj1}) and the burger equation (\ref{bugy}) leads to the differential equation (\ref{dif11}).  
In this section, we show that solving the quantum Hamiltonian-Jacobi through the burger equation (\ref{bugy}) leads to Kirchhoff equations (\ref{ufiu5}). These Kirchhoff equations (\ref{ufiu5}) are solved using Stieltjes electrostatic model. Hence, from theorem $1$ this model allows a Lagrangian formalism. We show that   when Stieltjes electrostatic model is used for solving quantum Hamiltonian-Jacobi it leads to quantum action $S$.

Theorem $8$ :  \textit{The quantum action or quantum characteristic function $S$, defined in  (\ref{ac}), is  }
\begin{eqnarray}
S= i\sum_{i\neq j}^nln\vert(x_k-x_j)\vert +i\sum_j\int\mathcal{W}(x_j) dx_j.\label{ac1}
\end{eqnarray}

Proof: We refer readers for proof \cite{iopj}. End of proof
We start by defining the quantum momentum function $p$ to be
\begin{equation}
p=\sum_{1\leq j\leq n,j\neq k}^n\frac{-i}{x_k-x_j}+i\mathcal{W}(x_j) .\label{ufiu55}
\end{equation}
This is justified because the quantum momentum function $p$ is the solution to the Riccati equation (\ref{qhj1}) and Burger equation (\ref{bugy}). The equation (\ref{ufiu55}) is complex pole evolution of Burger equation (\ref{bugy}) whose  pole structure is identical to quantum momentum function (\ref{uf0}). 
Substituting  (\ref{ufiu55}) in equation (\ref{acc}) and taking measure $dx=\prod_j dx_j$ \cite{iopj}, we get that equation  (\ref{ac1}).
 Using the definition of the quantum momentum function (\ref{mp}) and  minimizing it 
\begin{eqnarray}
p=\frac{dS}{dx_j}=0 \label{avo}
\end{eqnarray} gives (\ref{ufiu5}).
 Thus, equation (\ref{avo}) is minimizing quantum action $S$.  Therefore, the equation (\ref{ac1}) is quantum action.

In the classical Hamilton-Jacobi, equation of motion is governed by the Hamiltonian and the problem is solved by continuously transforming the Hamiltonian from the initial state to the final state through a series of canonical transformations. In random matrix theory, the probability distribution function (\ref{pdfo}) is a stationary solution of Fokker-Planck equation. From theorem $3$ and Corollary $1$ it is evident that, the Fokker-Planck equation is mapped to Schr\"{o}dinger equation by transformations (\ref{uio}). The exponential in transformation (\ref{uio}) is a Cole-Hopf transformations. Then, it is obvious that by a series of Cole-Hopf transformations leads to Burger equation whose complex pole evolution is given by Kirchhoff equations (\ref{ufiu5}). Integrating these Kirchhoff equations (\ref{ufiu5}) gives a random matrix (\ref{hamp}). Thus, this picture is exactly identical to classical Hamilton-Jacobi. Therefore, the $H(x)$ is the quantum action.

Theorem $9$ :  \textit{The quantum action (\ref{aci}) is the random matrix and the associated probability distribution function is the famous Cole-Hopf transformation. } \cite{kvs1}

Proof: We start with the random matrix probability distribution function  (\ref{pdfo}) taking $\beta=-i$ which reduces to Cole-Hopf transformation (\ref{ac})
given by
\begin{equation}\label{pdfoo}
\psi={\cal P}(x_1,\ldots,x_N)
= C_ne^{i H}
\end{equation}
where $H$ is given by equation (\ref{hamp}). Substituting (\ref{pdfoo}) in Schr\"odinger equation  (\ref{sc}) gives (\ref{qhj1}) with the quantum momentum function defined as $p=\frac{\partial H}{\partial x_j}$. If $V(x_j)=\int\mathcal{W}(x_j) dx_j $ is superpotential in  the random matrix $H$ is given by equation (\ref{hamp}) then it reduces to quantum action (\ref{acc}).
It is also evident from the theorem $2$ and $3$  that the random matrix probability distribution function (\ref{pdfo}) is the solution to the self-adjoint operator that is the Schr\"{o}dinger equation. This ends the proof.

Theorem $10$ :  \textit{If $V(x_j)=\int\mathcal{W}(x_j) dx_j $ is the superpotential then the  random matrix probability distribution (\ref{pdfo}) reduces to (\ref{pdf1}) for $\beta=2$, that is Gaussian unitary ensemble, in terms of the weight function and is of the form} \cite{kvs1}
\begin{equation}
P(x_o,x_1.....x_n)= \sum_i \vert \psi(x_i)\vert^2  \label{qpc}
\end{equation} 
\textit{the quantum mechanics probability distribution is in natural units.}

Proof:  It is clear from theorem $9$ that the random matrix probability distribution function   (\ref{pdfoo}) is Cole-Hopf transformation if  $V(x_j)=\int\mathcal{W}(x_j) dx_j $ is the superpotential in the equation (\ref{hamp}). In quantum mechanics, wave function for polynomial solutions are of the form \textit{square root of weight function times the polynomial}. Therefore, quantum mechanics probability distribution in natural units (\ref{qpc}) reduces to  random matrix probability distribution  (\ref{pdf1}) for $\beta=2$, that is the Gaussian unitary ensemble. The Vandermonde determinant (\ref{pdf1}) is written in terms of  polynomials \cite{ratm}. The weight function and superpotential are related as 
\begin{eqnarray}
{\cal W}(x_j)&=&\frac{d}{dx_j}ln(w(x_j)) \label{wio}
\end{eqnarray}
where $w(x_j)$ is the weight function. 
In quantum Hamilton-Jacobi the superpotential essentially captures the fixed poles, these are computed from the turning points of the classical potential, which are reflected  in the range of classical orthogonal polynomials in terms of weight function. 
This ends the proof.

It should be noted that from Corollary $1$ we have $\beta=-2i$. Then, from theorem $9$ and $10$ we have  $\beta=i$ and $\beta=2$ respectively. Suppose, if we use $\beta=-2i$ in theorem $9$ and rewrite potential in terms of weight function (\ref{wio}) we get random matrix probability distribution  (\ref{pdf1}) is consistent with Theorem $10$.  Probability distribution function in quantum mechanics (\ref{qpc}) ground state starts from $x_0$ and in the random matrix probability distribution function $\mathcal{P}(x_1.....x_n)$ starts from $x_1$ by definition.

\section{Illustrations}

As a first illustration, we consider the 1-D harmonic oscillator. The harmonic oscillator potential in natural units is  $V(x_j)=\frac 1 2 x_j^2$ such that the superpotential is $\mathcal{W}(x_j)=x_j$ \cite{kharebook}. Hence, Kirchhoff equation (\ref{ufiu5}) read as
\begin{equation}
\sum_{1\leq j\leq n,j\neq k}^n\frac{-i}{x_k-x_j}+ix_j=0.\label{1h}
\end{equation}
Integrating these equations  gives the quantum action from theorem $9$ and the wave function (\ref{pdfoo}) reduces to  Dyson-gas model (\ref{tt}) with $\beta=i\hbar$ in natural units $\hbar=1$.  The solution to 1-D harmonic oscillator are Hermite polynomials given by
\begin{equation}
\psi (x)=c_n e^{-{\frac {x^{2}}{2}}}\cdot H_{n}\left(x\right),\qquad n=0,1,2,\ldots .\label{hor1u}
\end{equation}
It is well known in literature that the equation (\ref{hor1u}) can be written as 
\begin{eqnarray}\label{hor1}
\psi (x)=c_n e^{-{\frac {x^{2}}{2}}}\cdot \begin{vmatrix}
H_0(x_0) & H_0(x_1) & \hdots  & H_0(x_n) \\ 
H_0(x_0) & H_1(x_1) &  & \vdots \\ 
\vdots & \;   \qquad   & \ddots  & \vdots   \\    
H_0(x_0) & \hdots & \hdots & H_n(x_n)
\end{vmatrix} .
\end{eqnarray}
The matrix in the equation (\ref{hor1}) is the Vandermonde determinant and its connection to polynomials is given reference \cite{ratm}. 
The probability distribution function in quantum mechanics is given by

Hence, then the probability distribution explicitly 
\begin{equation}
P(x_o,x_1.....x_n)= C_n e^{-X^2}\prod_{0\leq i\leq j\leq n}|X_i-X_j|^2\label{opi}
\end{equation}
Readers should note that we have used $X$  instead of $x$, it is well known in quantum mechanics that one obtains Hermite differential equation from Schr\"odinger equation for a dimensionless variables $X={\sqrt {\frac {m\omega }{\hbar }}}x$ as we have solved in natural units $m=\omega=\hbar=1$. Thus the quantum bound state problems mimic the Coulomb Dyson-gas model. It is clear from this example that weight function and the superpotential are related.

As another illustration, we consider the Coulomb problem and the wave function is given by
\begin{equation}
\psi(r)= r^{l+1}exp[-\frac{1}{2}r] L^{2l+1}_n(r)
\end{equation}
where the weight function $w_\beta(r_j)^\frac{1}{2}=r^{l+1}exp[-\frac{1}{2}r] $ and the $L^{2l+1}_n(r)$ are Laguerre polynomials.
We obtain the superpotential the Coulomb problem as follows 
\begin{eqnarray}
{\cal W}(r_j)&=&\frac{d}{dr_j}ln(w(r_j))=\frac{d}{dr_j}ln(r_j^{l+1}exp[-\frac{1}{2}r_j])=\frac{d}{dr_j}((l+1)lnr_j-\frac{1}{2}r_j)=\frac{(l+1)}{r_j}-\frac{1}{2}.
\end{eqnarray}
is the desired result.
As the third illustration, 
the weight function for the  Jacobi polynomials is given by $w(x_j)=(1-x_j)^\alpha(1+x_j)^\beta$, then the 
superpotential is given as follows 
\begin{eqnarray}
{\cal W}(x_j)&=&\frac{d}{dx_j}ln(w(x_j))=\frac{1}{w(x_j)}\frac{d}{dx_j}(w(x_j))=\frac{1}{w(x_j)}\frac{d}{dx_j}([(1-x_j)^\alpha(1+x_j)^\beta])
=\frac{\beta}{(1+x_j)}-\frac{\alpha}{(1-x_j)}
\end{eqnarray}
is the desired result. Hence, we have proved that the random matrix probability distribution function is the famous Cole-Hopf transformation and the underlying complex pole evolution of the Schr\"odinger equation is described quantum action in terms of random matrix. 
\section{Conclusion}
In this paper, starting with the quantum Hamilton-Jacobi approach, we have shown that the underlying complex pole evolution of the Schr\"odinger equation is described quantum action in terms of random matrix.  The random matrix probability distribution function is the famous Cole-Hopf transformation. The quantum bound state problems mimic the Coulomb Dyson-gas model. Thus the physical systems which are modeled using random matrices in terms of the Coulomb Dyson-gas model can use all quantum bound state potentials. Random matrix is the heart of quantum transformation theory.
\section*{Acknowledgments}
KVSSC acknowledges the Department of Science and Technology, Govt of India (Matrices scheme (D. O. No: MTR/2018/001046)) for financial support. Authors acknowledges Bindu A Bambah for her encouragement and comments.

\section{Appendix}
\subsection{ Dirac quantum characteristic function and quantum Hamilton-Jacobi }
When the Hamiltonian does not explicitly depend on time, the Hamilton's principle in classical Hamilton-Jacobi is given by
\begin{equation}
\mathcal{S}(x,t)=S(x)-Et\label{hp}
\end{equation} 
where $S(x)$ does not depend on $t$ and the Hamilton's characteristic function and action is defined as 
\begin{equation}
S=\int p dx \label{acc}
\end{equation}
where $p=\frac{\partial S}{\partial x}$ is the momentum function. For more details refer ro \cite{cs}.
substituting
\begin{equation}
\psi=exp(\frac{i}{\hbar}\mathcal{S}(x,t))\label{aci}
\end{equation} 
in the time dependent Schr\"odinger equation
\begin{equation}
i\hbar {\frac {\partial }{\partial t}}\psi (x,t)=\left[-{\frac {\hbar ^{2}}{2m}}{\frac {\partial ^{2}}{\partial x^{2}}}+V(x,t)\right]\psi (x,t)\,.
\end{equation}
gives
\begin{eqnarray}
\frac{\hbar}{i}\frac {\partial ^{2}}{\partial x^{2}}\mathcal{S}(x,t)+ \left(\frac{\partial}{\partial x} \mathcal{S}(x,t)\right)^2=2m\left(-\frac{\partial }{\partial t}\mathcal{S}(x,t)-V(x)\right).\label{qhji}
\end{eqnarray}
Substituting equation (\ref{aci}) in equation (\ref{qhji}) reduces to (\ref{qhj0}). Thus, $S$ in equation (\ref{qhj0}) is the Dirac quantum characteristic function.
\subsection{Supersymmetry}
In supersymmetric quantum mechanics, the superpotential $\mathcal{W}(x)$ is defined in terms of the intertwining operators $ \hat{A}$ and $\hat{A}^{\dagger}$  as
\begin{equation}
\hat{A} = \frac{d}{dx} + \mathcal{W}(x), \qquad \hat{A}^{\dagger} = - \frac{d}{dx} + \mathcal{W}(x), 
\label{eq:A}
\end{equation}
This allows one to define a pair of factorized Hamiltonians $H^{\pm}$ as
\begin{eqnarray}
H^{+} &=&     \hat{A}^{\dagger} \hat{A}     = - \frac{d^2}{dx^2} + \mathcal{V}^{+}(x) - E, \label{vp}\\
H^{-} &=&     \hat{A}  {\hat A}^{\dagger}     = - \frac{d^2}{dx^2} + \mathcal{V}^{-}(x) - E, \label{vm}
\end{eqnarray}
where $E$ is the factorization energy. 
The partner potentials $\mathcal{V}^{\pm}(x)$ are related to $\mathcal{W}(x)$ by 
\begin{equation}\label{gh}
\mathcal{V}^{\pm}(x) = \mathcal{W}^2(x) \mp \mathcal{W}'(x) + E.
\end{equation}


\begin{thebibliography}{99}
	\bibitem{ws}A. Sommerfeld, Sitzungsber. Bayr. Akad. , p. 425 (1915); W.
	Wilson, Phil. Mag. 31, 156 (1916);M. Born, The Mechanics
	of the Atom (G. Bell, London, 1927), translation by J. W.
	Fisher
\bibitem{jor} P. Jordan, Z. Phys. 38, 513 (1926); 40, 809 (1927); 44, 1 (1927).
\bibitem {dir} P. A. M. Dirac, Proc. R. Soc. London 113A, 621 (1927).
\bibitem {dir1}P. A. M. Dirac, Phys. Z. Sowjetunion 3, 64 (1927); Rev. Mod.
Phys. 17, 195 (1945).
\bibitem{dir2} P. A. M. Dirac, The Principles of Quantum Mechanics (Oxford
University Press, London, 1958), fourth edition.
\bibitem{fey} R. P. Feynman, Rev. Mod. Phys. 20, 367 (1948); also R. P.
Feynman and A. R. Hibbs, Quantum Mechanics and Path In
tegrals (McGraw-Hill, New York, 1965).
sJ. Schwinger, Quantum Electrodynamics (Dover, New York,

1958); this reprint volume contains papers relevant to our dis-
cussion, including the first paper cited in Ref. 3, and the pa-
pers cited in Refs. 5 and 8.
\bibitem{schw}J. Schwinger, Quantum Electrodynamics (Dover, New York,

1958); this reprint volume contains papers relevant to our dis-
cussion, including the first paper cited in Ref. 3, and the papers cited in Refs. 5 and 8.

\bibitem{fin} R. J. Finkelstein, Nonrelati Uistic Mechanics (Benjamin-
Cummings, CA 1973).

\bibitem {schw1} J. Schwinger, Quantum ICinematics and Dynamics (Benjamin-
Cummings, CA 1970), and references therein. (1953)
\bibitem{qhj} R. A Leacock and M. J. Padgett  Phys. Rev. Lett. 50, 3, (1983) 
\bibitem{qhj1} R. A. Leacock and M. J. Padgett Phys. Rev. D28, 2491, (1983).
\bibitem{zo}E.Gozzi, Physics Letters B, 158, 6, 489, (1985)

\bibitem{chj} Sirin A. Buyukasik , Oktay K. Pashaev 	arXiv:1005.5059 [nlin.SI]

\bibitem{kvs1} K.V.S Shiv Chaitanya,  Foundations of Physics, \textbf{49}, 351, (2019)
\bibitem{kvs} K. V. S. Shiv Chaitanya,  PRAMANA journal of physics, Vol. 83, No. 1, 139, (2014). 
\bibitem{st} T.J. Stieltjes, Sur Quelques theorems d'algebre, Comptes Rendus de 
l'Academie des
Sciences, Paris, 100 (1885), 439-440; Oeuvres Completes, Vol. 1, 440-441.
\bibitem{st1} T.J. Stieltjes, Sur Quelques theorems d'algebre, Comptes Rendus de l'Academie des
Sciences, Paris, 100 (1885), 620-622; Oeuvres Completes, Vol. 1, 442-444.

 
\bibitem{met} Mehta, M.L. (2004). Random Matrices, 3rd Edition, Pure and Applied Mathematics
(Amsterdam), 142, Amsterdam, Netherlands: Elsevier/Academic Press.
\bibitem{wig} Wigner, E. P., Gatlinburg Conference on Neutron Physics. Report 2309:59, Oak
Ridge National Laboratory, Oak Ridge, TN (1957).\bibitem{qcd}J.J.M. Verbaarschot, T. Wettig, Ann.Rev.Nucl.Part.Sci. 50, 343-410, (2000).
\bibitem{qtqd}Y. Alhassid, Rev.Mod.Phys.72:895-968,2000.

\bibitem{sas} B Sriram Shastry and Abhishek Dhar, J. Phys. A: Math. Gen. 34, 6197, (2001).


\bibitem{cs}Peter J. Forrester, MSJ Memoirs
Volume 1,  97-181, (1998).
\bibitem{phi} Philip Choquord and Stephane Rey, \textit{European Journal of Physics}, \textbf{17} 2, 45 (1996)
\bibitem{tar} Tarun R Tummaru \textit{et al} \textit{Physics PDFs A}\textbf{381},47,3917(2017)
 \bibitem{zaf} Zaffar Ahmed and Sudhir R. Jain, 2000, Pramana 54,3, 413
 \bibitem{guy} Guy Auberson \textit{et al} 2001, J Phys A: Mathematical and general 34, 4, 695.
 \bibitem{sud} Sudhir R Jain and Avinash Khare, 1999, Physics PDFs A, 262, 1, 35.
 
 \bibitem{dan} Daniel Alenso and Sudheer R Jain, 1996, Physics PDFs B, 387, 4, 812.
 \bibitem{rm} Alan J. Izenman, Introduction to Random-Matrix Theory, lecture notes.
\bibitem{ran1} Greg W. Anderson, Alice Guionnet, and Ofer Zeitouni, \textit{An Introduction to Random Matrice}s, volume 118 of Cambridge studies in advanced mathematics. Cambridge University Press, 2009.
\bibitem{ran2} Jinho Baik, Percy Deift, and Kurt Johansson. On the distribution of the length of the longest increasing subsequence of random permutations. Journal of the American Mathematical Society, 12:1119–1178, 1999.
\bibitem{ran3}Percy Drift. Orthogonal Polynomials and Random Matrices: A Riemann-Hilbert Approach, volume 3 of Courant Lecture Notes in Mathematics. Courant Institute of Mathematical Sciences, 1999.

\bibitem{ran4}Sudhir R. Jain, Czechoslovak Journal of Physics, 56, 9, 1021, (2006) between eigenvalues of a random matrix. Comm. Math. Phys. 19 (1970), no. 3, 235--250
\bibitem{dyson}Freeman J. Dyson, 1972, J Math Phys \textbf{13} 90.
\bibitem{sz} G. Szeg\H{o}, Orthogonal Polynomials, Fourth Edition, Amer. Math. Soc., Providence, 1975.
\bibitem{nelson} E Nelson, Phy Rev, Vol 150, No 4, 1079, (1966).
\bibitem{sree} S. Sree Ranjani, K. G. Geojo, A. K. Kapoor, P. K. Panigrahi, Mod. Phys. Lett. A. Vol 19, No. 19,  1457, (2004).
\bibitem{inc}E. L. Ince, Ordinary Differential equation (Dover Publication Inc, New York 1956).
\bibitem{chood}Choodnovsky, D. V., Choodnovsky, G. V, Nuovo Cimento B 40:339-53 (1977)

\bibitem{kharebook} F. Cooper, A. Khare, U. P. Sukhatme, 2001 {\it Supersymmetric quantum mechanics} (Singapore: World Scientific Publishing Co. Ltd.)
\bibitem{ratm}Giacomo Livan, Marcel Novaes, Pierpaolo Vivo, "Introduction to Random Matrices Theory and Practice" (Springer 2017)
\bibitem{cs}  Herbert Goldstein , Charles Poole Jr. , John Safko, "Classical Mechanics",  Pearson; 3rd edition (26 June 2001). 

\bibitem{iopj} Giacomo Livan Marcel NovaesPierpaolo Vivo, "Introduction to Random Matrices Theory and Practice",  SpringerBriefs in Mathematical Physics book series (BRIEFSMAPHY, volume 26).
\end{thebibliography}
\end{document}